\documentclass[preprint,12pt]{elsarticle}

\usepackage{amssymb}

\usepackage[nodots]{numcompress}

\usepackage{subfigure}

\journal{Nuclear Instruments and Methods A}

\begin{document}

\begin{frontmatter}



\title{A Highly Efficient Neutron Veto for Dark Matter Experiments}


\author{Alex Wright\corref{cor1}}
\cortext[cor1]{Corresponding author: Department of Physics, Jadwin Hall, Princeton University,
  Princeton NJ, 08540, USA. Tel: (609) 258-0154 Email: ajw@princeton.edu}
\author{Pablo Mosteiro}
\author{Ben Loer}
\author{Frank Calaprice}

\address{Physics Department, Princeton University, Princeton NJ,
  08544, USA}

\begin{abstract}

We present a conceptual design for an active neutron veto, based on
boron-loaded liquid 
scintillator, for use in direct-detection dark matter experiments. The
simulated efficiency of a 1 meter thick veto, after including the
effects of neutron captures in the inner detector
and inefficiencies due to feed-throughs into the veto, is greater than
99.5\% for background events produced by radiogenic neutrons, while
the background due to externally produced cosmogenic neutrons is reduced by more than
95\%. The ability of the veto to both significantly suppress, and
provide {\it in situ} measurements of, these two dominant
sources of background would make the next generation of dark matter
experiments much more robust, and dramatically improve the credibility
of a dark matter detection claim based on the observation of a few
recoil events. The veto would also allow direct extrapolation between
the background-free operation of a small detector and the physics
reach of a larger detector of similar construction.

\end{abstract}


\end{frontmatter}


\section{Introduction}
\label{intro}

As the sensitivity of direct detection dark matter experiments
continues to improve, the suppression of neutron backgrounds through
the selection of highly radiopure materials and passive shielding
becomes much more difficult. Indeed, experiments are reaching
sensitivities where backgrounds estimates using calculations based on
{\it ex situ} measurements of representative material samples are no
longer sufficiently robust, as relatively modest differences in production methods
and material handling procedures can have a significant effects on the
overall background in extremely pure materials. Therefore, the ability to precisely measure background
levels {\it in situ} will be critical to giving future dark matter
experiments the ability to make credible claims of dark matter
detection.

As direct detection dark matter experiments seek to detect WIMP
dark matter by observing nuclear recoils produced by WIMP interactions
with nuclei, and as single scatter neutron events can produce nuclear
recoils that are indistinguishable from WIMP interactions, the issue
of {\it in situ} background control is particularly relevant in
considering neutron backgrounds. These neutrons are produced both within
the material comprising the detectors themselves and in the materials
surrounding them through
radiogenic processes (($\alpha$, n) reactions and spontaneous fission)
and by spallation by cosmic-ray muons. The neutron-induced background
in dark matter detectors has typically been minimized by constructing the detectors
from
materials containing extremely low levels of radioactive contaminants,
by passively shielding the experiments against externally produced
radiogenic neutrons, and by operating the experiments in underground
laboratories where the cosmogenic neutron rate is
reduced. As mentioned above, however, these strategies, are reaching their practical
limits; the size and sensitivity of dark
matter experiments is reaching the point where it is extremely
difficult to demonstrate, using {\it ex-situ} assay techniques, that the
materials used in the construction of a given detector are
sufficiently free from radioactivity that the production of a small
number of neutron-induced recoil events is unlikely. In addition, only the deepest of underground laboratories have
cosmogenic neutron fluxes low enough to permit the operation of
next generation of dark matter experiments in conventional
shielding with a robust expectation having no cosmogenically-induced
background. These effects combine to both limit the sensitivity of
experiments and to make it extremely difficult for any experiment to
claim, with high confidence, that the detection of a few recoil
events constitutes a detection of dark matter. Indeed, at the present
time all of the leading dark matter
experiments see some nuclear-recoil-like events, so these experiments
are either background limited or unable to make a detection claim based
on the few events observed.

Replacing the passive neutron shielding with an active neutron veto is one
way in which experiments can both lower their neutron backgrounds and
make precise {\it in situ} assays of this important background. A
high-efficiency active neutron veto would not only significantly 
improve the sensitivity of an experiment which would otherwise be
limited by neutron backgrounds, but it would also allow an experiment
which has been successful in achieving a low neutron background (and
which hence has very few observed coincidences between nuclear
recoils in the dark matter detector and signals in the veto) to
demonstrate convincingly that the number of neutron-induced recoil
backgrounds possible in the data is extremely small. It is conceivable
that with a high efficiency veto these limits could be as
low as a few hundredths of an event. This would help to give experiments
the ability to make convincing claims of dark matter detection based
on the observation of a few events, and help to move the field from
working to set better limits on the dark matter interaction cross
section to attempting to detect dark matter interactions. In
addition, the ability to extrapolate a carefully understood
background, coupled with the direct suppression afforded by the
neutron veto, would allow a smaller experiment to
demonstrate directly the potential for a much larger experiment of
similar construction to operate background free. This extrapolation
ability would be extremely helpful in guiding the development of ever
larger and more sensitive detectors. The practical utility of all of
these applications, however, is dictated by the absolute efficiency of
the neutron veto; as a result, the development
of a highly efficient veto system is extremely desirable.  

The obvious benefit of neutron veto capability has led most current
dark matter experiments to implement veto procedures. In some, the veto
is achieved by segmenting (through position reconstruction
and/or physical segmentation) the active volume of the dark matter
detector to look for neutrons that produce more than one recoil
event. Other experiments [\cite{warp_veto}, \cite{zeplin_veto}] have
deployed separate, dedicated veto systems. In the future, very large,
monolithic detectors could veto neutron events quite efficiently using
internal coincidences, except near the
detector walls where there is a reasonable probability for the
recoiling neutron to escape, and where neutrons with only enough
energy to produce a single detectable recoil are most likely to
interact. In order to have a highly efficient neutron veto, then,
these large detectors will have to either take a reasonably large cut in fiducial volume
or install an external neutron veto.

A neutron veto system with a very high detection efficiency can be
produced by surrounding a dark matter detector with a layer of liquid
scintillator. Such a veto, with a thickness of order 1 meter, is sufficient to detect a very high percentage of
the radiogenic neutrons produced by the inner detector, and a
significant fraction of cosmogenic neutrons. However, at about 250
$\mu$s, the
capture time for thermal neutrons in a liquid scintillator is rather
long. This means that in order to efficiently veto
the (promptly produced) neutron-induced dark matter backgrounds, veto
windows of millisecond duration are necessary. With such a long veto
window, the background rate in the scintillator must be quite low,
less than $\sim$100 Hz, to keep the veto-induced dead time in the
dark matter detector from becoming significant. Achieving such a low
event rate in the veto requires bulky
and expensive passive shielding; even the use of conventional
photomultiplier tubes to instrument the veto becomes difficult, as the
event rate due to radioactive contaminants in the PMTs
themselves would be excessive.

In this paper we show that relatively compact, highly efficient
neutron vetoes for
dark matter detectors are practically realizable through the use of
boron-loaded liquid scintillator. We have studied such a veto using
Geant4-based Monte Carlo simulations in the context of the proposed
DarkSide-50 dark matter detector.  The efficiency of the veto is shown
to be very high, even after taking into account neutrons which do not
escape the inner detector and potential inefficiencies due to feed-through
penetrations. The total event rate in the veto from internal and
external sources is estimated and found (with a relatively modest
amount of passive shielding against external gamma-rays) to be
acceptable, even using conventional PMTs. In large part, this tolerance
of veto rate is due to
the decrease in the neutron capture time afforded by the large capture
cross-section of $^{10}$B. Optical simulations of the veto show that
the very low energy threshold necessary to reliably detect the
reaction products of neutron capture on $^{10}$B is comfortably
achievable, even under fairly conservative assumptions about the
optical characteristics of the veto.

\section{Boron-Loaded Scintillator}
\label{scint}

Neutron detection using boron-loaded liquid scintillator, produced by
adding tri-methyl borate (TMB) to standard
scintillator cocktails, was first
investigated in \cite{Muehlhause_and_thomas} and subsequently
developed into practical detectors by \cite{Bollinger_and_thomas}. 

$^{10}$B, which has a natural abundance of about 20\%, captures thermal
neutrons with a very high (3837(9)b) total capture cross section via
two channels \cite{endf}: 
\begin{eqnarray*}
^{10}\textrm{B} + n &\rightarrow& ^7\textrm{Li (g.s.)}+\alpha\textrm{
    \phantom{aaaaaaaaaaaaaaaaaaaaaaaaaaaaa} 6.4\%}\\
&\rightarrow& ^7\textrm{Li$^*$} + \alpha \textrm{, } ^7\textrm{Li$^*$}
  \rightarrow ^7\textrm{Li} + \gamma
  \textrm{(478 keV) \phantom{aaaaaaaala} 93.7\%}
\end{eqnarray*}
Importantly for the current discussion, the nuclear recoil reaction
products carry a significant amount of energy (in the decay to the
excited state of $^7$Li, E($\alpha$) = 1471 keV and E($^7$Li) = 839
keV, while for the ground state decay E($\alpha$) = 1775 keV and
E($^7$Li) = 1015 keV). The light output from nuclear recoils in liquid
scintillator is heavily quenched, to the level of 50-60 keV$_{ee}$ \cite{Greenwood_and_chellow, wang_et_al}, but, as will be shown, this is still
detectable. The ability to detect neutron captures via scintillation
from the nuclear recoil products makes relatively compact neutron detectors
made from boron-loaded scintillator possible, as it is not necessary
for the detector to be large enough to contain capture-induced gamma-rays. 

Many applications of boron-loaded liquid scintillator neutron detectors
make use of the delayed coincidence signals produced when energetic
neutrons first produce recoil proton events during thermalization and
are subsequently captured by $^{10}$B. This is the basis of the
``capture-gated neutron spectrometer'' technique \cite{knoll}, and can
be used to produce neutron detectors that can be operated in
relatively high background environments\footnote{Where the amount of
light collected is high, gamma-ray interactions and neutron capture events in
boron-loaded scintillator can also be distinguished using
pulse shape analysis \cite{wang_et_al}.}. Development of neutron
detectors based on this technique is ongoing (see, e.g.,
\cite{swiderski}), particularly for national security applications. In high
efficiency neutron detectors, like the neutron vetoes described here, 
the coincidence technique cannot be exploited due to its lower overall
efficiency (in the DarkSide-50-based
simulations described below, only $\sim$80\% of radiogenic neutrons produced a coincidence event in the
boron-loaded scintillator).

TMB-loaded scintillators have also been investigated for use in large, low
background particle detectors, as reported in \cite{Borex_proposal,
  reines_scint}. In particular, scintillator cocktails with TMB
loading of up to 80\%, and with light output,
optical attenuation, radio-purity, and scintillator stability
properties suitable for a large, low-background neutrino experiment
were identified in \cite{Borex_proposal}.

\section{Monte Carlo Modeling and Validation}

The Monte Carlo studies which constitute the bulk of this report were
carried out using the Geant4 Toolkit (version 4.9.3) \cite{geant4},
via a flexible physics and geometry interface developed at Princeton University. 

The energy spectra used in simulating the radiogenic neutrons produced in
different materials were determined by first calculating the
individual spectra expected from ($\alpha,n$)
and fission processes due to $^{235}$U, $^{238}$U, and $^{232}$Th
chain activities in each material using the
SOURCES4A software package \cite{sources4}. These ``component
spectra'' were then combined, using measured
ratios of the different radioactive species, to give the total
radiogenic neutron spectrum for each detector material. Cosmogenic
neutrons were generated with the (depth dependent)
approximate energy spectrum described in \cite{Mei_and_Hime}.
 
In what follows, ``pure scintillator'' is pseudocumene
(C$_9$H$_{12}$, $\rho=0.876$g/cm$^3$), while ``boron
loaded scintillator'' is 50\% w/w TMB in pseudocumene
(this gives a composition of 62.3\% C, 23.1\% O, 9.4\% H, and 5.2\% B by
mass) and is assumed to have a density equal to that of pseudocumene. The
wavelength shifter necessary for the efficient optical performance of
the scintillator, which is likely to be 2,5-diphenyloxazole (PPO)
at the level of a few grams per litre in both boron-loaded and
unloaded scintillator, will
have a negligible effect on the overall neutron capture and is
neglected in the simulation. All
elements, including boron, are assumed to have natural isotopic abundances.

In order to confirm that the simulation provides a reasonable
reproduction of neutron behavior, a number of ``benchmark''
comparisons have been made:

\begin{enumerate}

\item The simulated mean capture time for radiogenic neutrons in
  pseudocumene is 253$\pm$1 $\mu$s. This can be compared to the
  256.0$\pm$0.4 $\mu$s neutron capture time observed by Borexino \cite{Borex_marco}, which uses the same pseudocumene scintillator
  simulated here.
\item In the simulation, 99.1\% of thermal neutrons captured in the
  boron-loaded scintillator were captured by $^{10}$B. Based on the
  ratio of the neutron capture cross-sections of the scintillator components (from
  \cite{jendl}), one expects that
  98.7$\pm$0.2\% of thermal neutrons should be captured by $^{10}$B
  (at 19.9\% $^{10}$B abundance). 
\item The fraction of neutron captures by $^{10}$B that produce
  $^7$Li in the first excited state is 93.67$\pm$0.07\% in the simulation, in
  good agreement with the expected value of 93.7\% \cite{endf}
\item The mean free paths of neutrons of different energies deduced
  from the results of the
  simulation, as well as the mean free paths expected based on the
  cross sections in \cite{endf} are shown in Table \ref{tab:mfps}. 
\end{enumerate}
These comparisons suggest that, although not perfect, neutron propagation and capture are
reproduced reasonably well by the simulation code, and that the predictions
of the neutron veto simulation might therefore be expected to provide
reasonable predictions of the performance of an actual veto.

\begin{table}[htbp]
\centering
\begin{tabular}{| c | c c | c c | c c |}
\hline
Neutron Energy & \multicolumn{2}{ |c|}{MFP in Water} &
\multicolumn{2}{|c|}{MFP in Pure} & \multicolumn{2}{|c|}{MFP in Loaded}\\ 
(MeV) &  \multicolumn{2}{|c|}{(cm)} &
\multicolumn{2}{|c|}{Scintillator (cm)} &
\multicolumn{2}{|c|}{Scintillator (cm)}\\
\hline
 & Sim. & XS & Sim. & XS & Sim. & XS \\ 
\hline
10 & 8.9 & 9.4 & 9.9 & 10.5 & 10.5 & 10.9\\
20 & 12.8 & 11.5 & 13.3 & 11.8 & 13.6 & 12.3\\
50 & 26.7 & 19.5 & 29.9 & 21.8 & 30.4 & -- \\
100 & 40.9 & 38.0 & 45.3 & 43.3 & 46.6 & -- \\
200 & 63.5 & -- & 70.9 & -- & 71.1 & -- \\
\hline
\end{tabular}
\caption{The mean free path (MFP) of neutrons of different energies in
  water, scintillator and boron-loaded scintillator, as simulated
  (``Sim.'') and as calculated using the neutron interaction cross
  sections in \cite{endf} and the chemical composition of the veto
  (``XS,'' where data is available). In the Monte Carlo, only those interactions resulting in the
  creation of a secondary particle with more than 1 eV of kinetic
  energy were directly recorded. Thus, some low energy interactions,
  low angle forward scattering for example, were excluded in the
  determination of the Monte Carlo MPFs; this could help to account for the longer
interaction lengths in the Monte Carlo.}
\label{tab:mfps}
\end{table}

\section{Neutron Detection with Boron-Loaded Scintillator}

\begin{figure}[htbp]
\centering
\subfigure[]{\scalebox{0.35}{\includegraphics{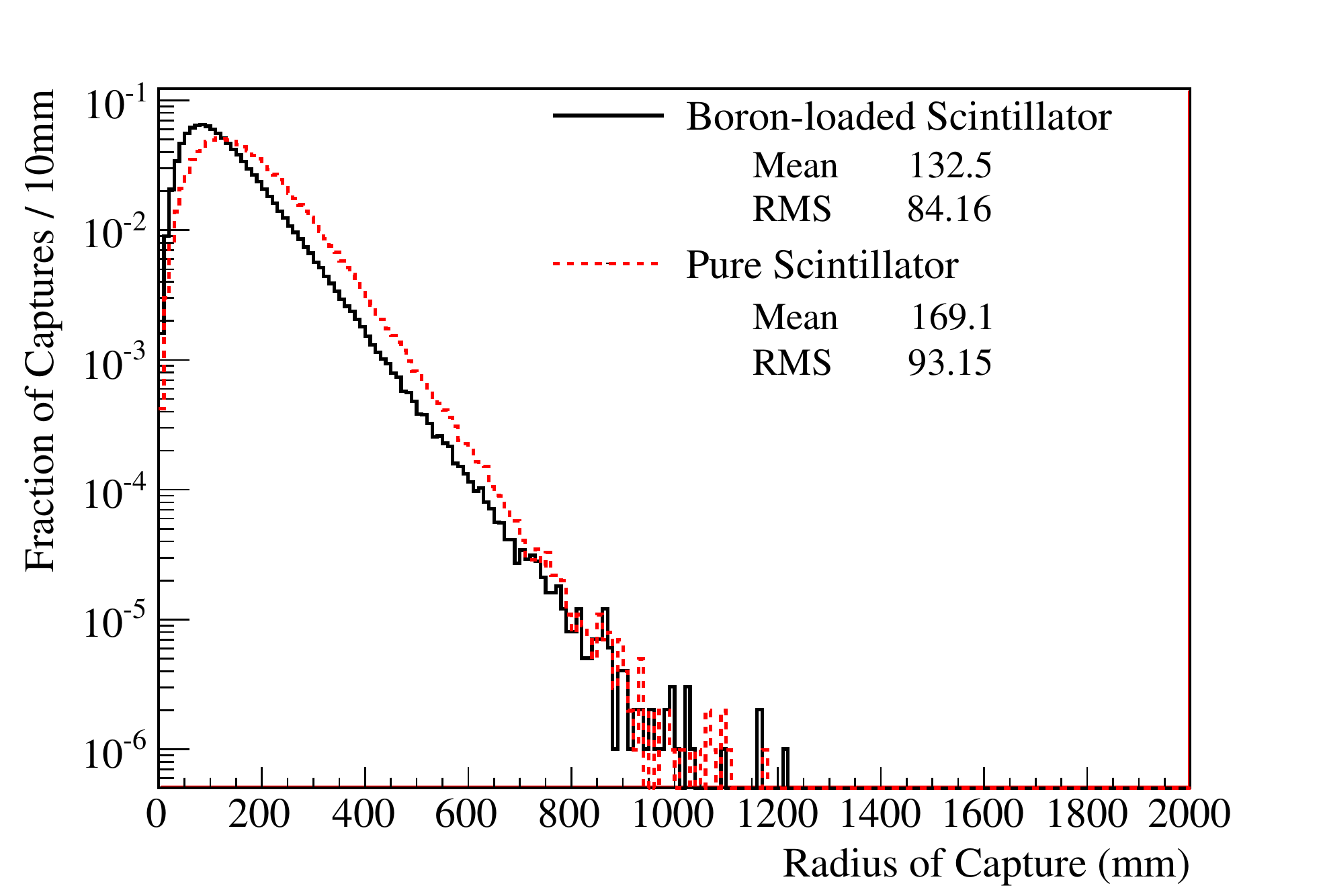}}}
\subfigure[]{\scalebox{0.35}{\includegraphics{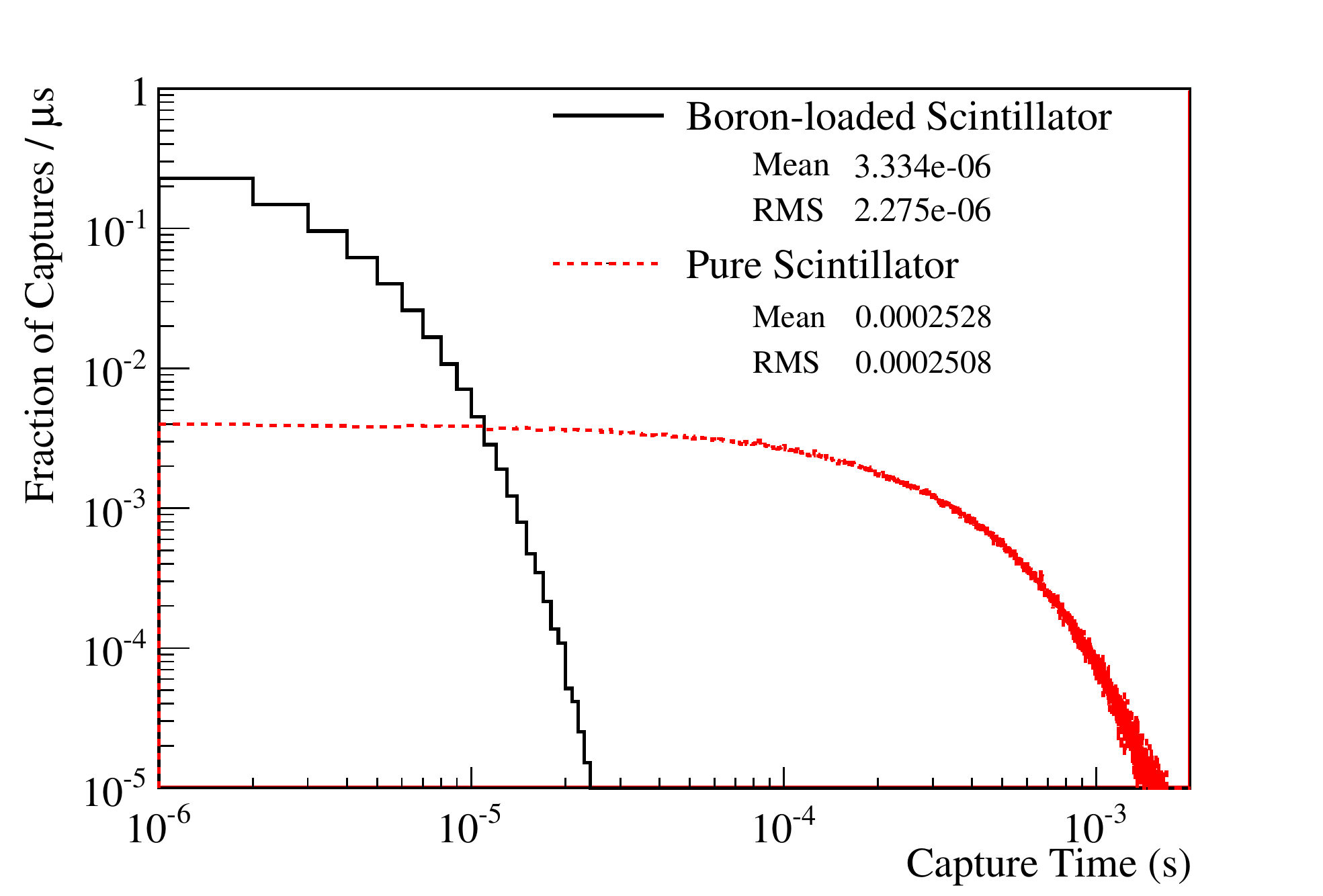}}}
\caption{The simulated distributions in radius and time of the captures of
  centrally produced radiogenic neutrons in large, uniform volumes of boron-loaded and pure
  liquid scintillator.}
\label{fig:rdist}
\end{figure}

Using the Monte Carlo described above, we have simulated the thermalization and capture of
radiogenic neutrons in pure and boron-loaded scintillators. Figure \ref{fig:rdist} shows the radial and time
distributions of the simulated neutron captures. As can be seen, the addition of the boron
reduces the mean neutron capture time by more than a factor of 100,
from 253 $\mu$s to 2.3 $\mu$s.

The distributions shown in Figure \ref{fig:rdist}, while illustrative,
do not correspond exactly to the distribution of veto event production in the
scintillator. Two main factors contribute to this
difference: first, capturing a neutron does not
necessarily result in a veto signal at the point of capture, especially in pure
scintillator. Instead, secondary particles, particularly gamma rays,
which propagate some distance before depositing
a detectable amount of energy in the scintillator, 
are often produced. Second, it
is possible that the neutron deposits sufficient energy via nuclear
recoils to produce a veto signal before being captured. As these recoils
occur promptly, the recoil-induced veto signals can occur
significantly earlier than the capture signals.

To investigate these effects, a ``veto signal'' was assumed to be
generated in the simulation if
 40 keV$_{ee}$ or more was deposited in the scintillator within any
1 $\mu$s time window. The quenching of heavy particles in the TMB-loaded
scintillator was assumed to be identical to quenching in unloaded
scintillator. Quenching for protons and alpha particles were treated
separately following \cite{proton_quench} and \cite{alpha_quench}; all
heavier recoils were quenched as carbon \cite{carbon_quench, proton_quench}. It might be expected that the use
of quenching values from undiluted scintillator would underestimate
the quenching effect in a diluted scintillator; however, the
pure scintillator quenching values give 50 keV$_{ee}$ energy deposition
in the scintillator from the recoil products of $n + ^{10}$B
$\rightarrow \alpha + ^7$Li$^*$, in good agreement with
observations in TMB-loaded (and hence diluted) scintillator \cite{Greenwood_and_chellow, wang_et_al}. 

\begin{figure}[htbp]
\centering
\subfigure[]{\scalebox{0.35}{\includegraphics{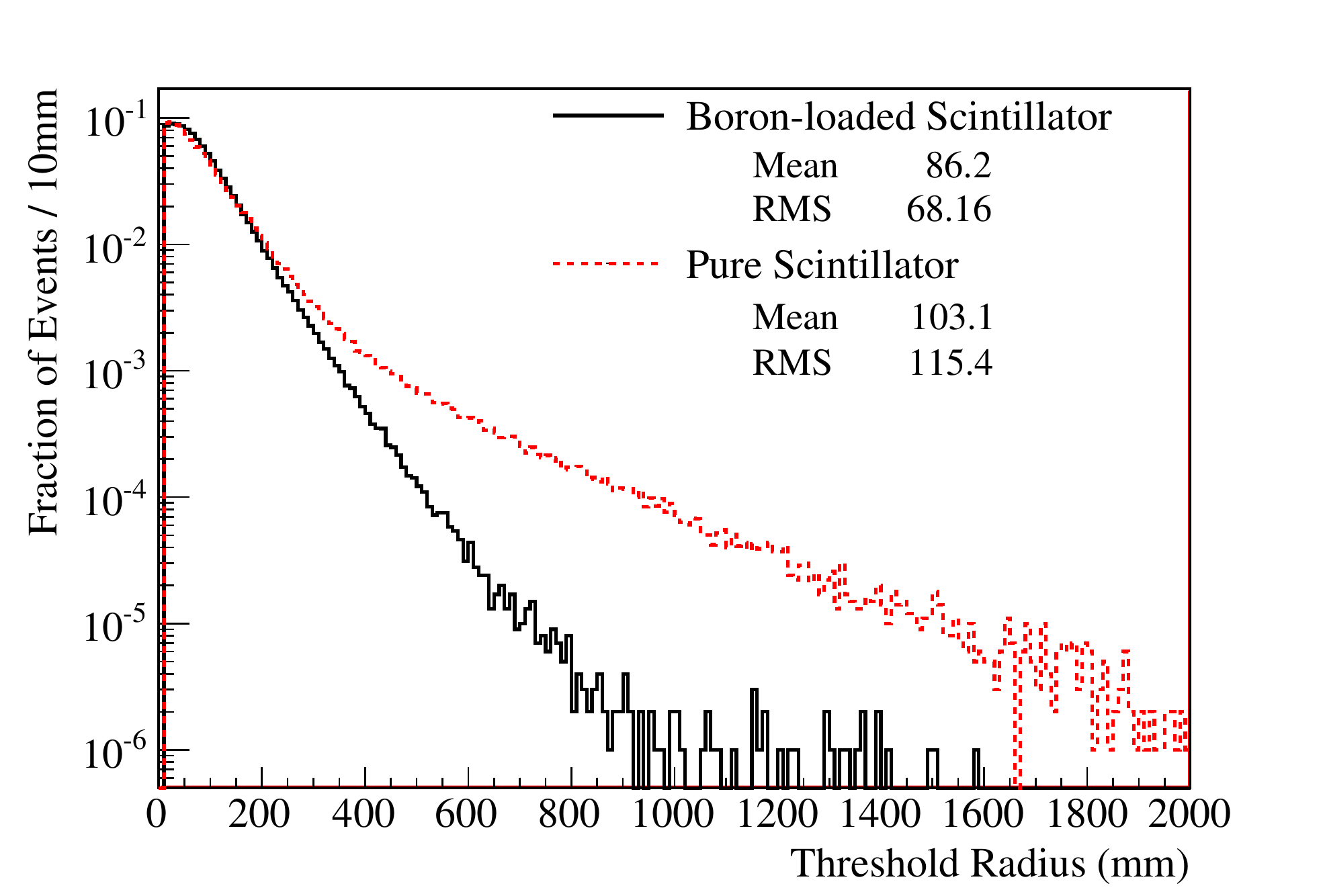}}}
\subfigure[]{\scalebox{0.35}{\includegraphics{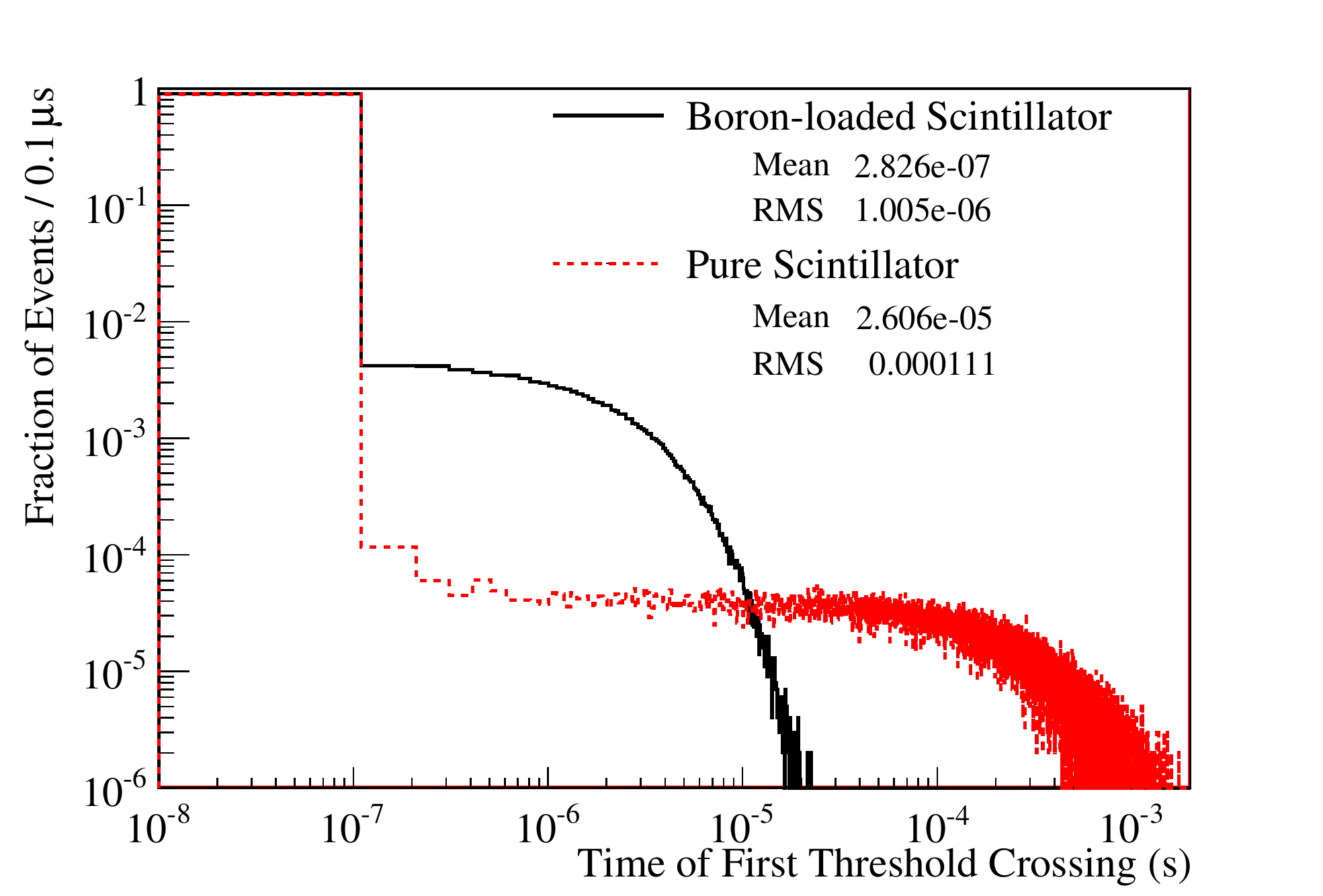}}}
\caption{The radial positions and times at which the first veto
  trigger associated with each neutron event was generated.The x-axis
  of (b) has been changed relative to Figure \ref{fig:rdist} to emphasize the
prompt veto events created by nuclear recoils during neutron
thermalization: such a prompt veto is produced by approximately 90\%
of neutrons in both pure and boron-loaded scintillator.}
\label{fig:thresh_dist}
\end{figure}

Figure \ref{fig:thresh_dist} shows the time distribution for the
production of the first veto signal (if any) for each neutron event, and the
detector radius necessary to contain the energy deposited in that trigger. As expected, in
pure scintillator the spatial
distribution of veto triggers is broader than the distribution of
neutron captures because of gamma-ray propagation. For boron-loaded
scintillator, by contrast, this broadening does not occur because the
recoil daughters from neutron capture on $^{10}$B deposit sufficient
energy at the site of capture to produce a veto signal. In fact, in
boron-loaded scintillator the distribution of veto trigger production
is narrower than the distribution of neutron captures, due to the
generation of veto triggers by neutrons scattering
prior to capture. Tables \ref{tab:eff_rad} and \ref{tab:eff_time}
contrast the radial and time windows necessary to contain, with
different probabilities, the veto signal after the production of a
radiogenic neutron in pure and boron-loaded scintillator. As can be
seen, the addition of boron decreases the windows in both radius and time
necessary to detect the veto neutrons with high probability.

\begin{table}[htbp]
\centering
\begin{tabular}{c c c}
\hline\hline
Containment Probability &  Pure Scintillator &
Loaded Scintillator\\
 & Radius (cm) & Radius (cm)\\
\hline
70\% & 11.1 & 10.2 \\
90\% & 21.0 & 17.2 \\
95\% & 29.1 & 21.7 \\
98\% & 44.8 & 28.0 \\
99\% & 60.4 & 32.9 \\
99.5\% & 78.0 & 38.1 \\
99.9\% & 129.7 & 51.6 \\
99.99\% & -- & 136.5 \\
\hline\hline
\end{tabular}
\caption{The radius required to contain the scintillator veto signal
  with different probabilities.}
\label{tab:eff_rad}
\end{table}

\begin{table}[htbp]
\centering
\begin{tabular}{c c c}
\hline\hline
Detection Efficiency & Time in Pure & Time in Loaded\\
&  Scintillator ($\mu$s) & Scintillator ($\mu$s)\\
\hline
70\% & 0.08 & 0.08 \\
90\% & 7.8 & 0.1 \\
95\% & 185 & 1.7\\
98\% & 421 & 3.8\\
99\% & 603 & 5.4 \\
99.5\% & 788 & 7.0\\
99.9\% & 1282 & 10.9\\
99.99\% & -- & 22.0\\
\hline\hline
\end{tabular}
\caption{The time interval after neutron production (and hence any prompt
  recoils in the WIMP detector) necessary to contain the veto signals with
  different probabilities.}
\label{tab:eff_time}
\end{table}

\subsection{Other Loading Options}

We note that the reduction in the average capture time of neutrons in
boron-loaded scintillator compared to pure scintillator could also be
achieved by loading the scintillator with other isotopes possessing
large neutron capture cross-sections, among them $^6$Li, $^{113}$Cd, and
$^{157}$Gd. Neutron capture on the latter isotopes is detected through the
emission of gamma-rays, which raises the prospect that, as with
neutron captures by protons, larger scintillator volumes would be
necessary to contain the gamma rays. In our simulations, however, we
find the neutron capture performance of a veto composed of scintillator loaded to
0.6\% by weight with natural gadolinium offers very similar
performance, in both capture time and radius of energy deposition, to the boron-loaded scintillator. The production of a
cascade of gamma-rays, rather than a single photon, by neutron
captures on $^{157}$Gd seems to be responsible for the improved spatial
performance of gadolinium relative to pure scintillator. Gadolinium
loading offers the advantage that the majority of neutron captures on
$^{157}$Gd deposit more energy in the veto than do background gamma
rays; this raises the possibility that neutron detection in Gd-loaded
scintillator might be more robust against $\gamma$-ray backgrounds. In simulations including the inner WIMP detector from
DarkSide-50 (described below), however, 6\% of neutron events
deposited less than 2 MeV$_{ee}$ in a veto composed of Gd-loaded scintillator and hence fall below the
onset of the background from external gamma-rays, and 0.5\% fell below
100 keV. Therefore, while gadolinium loading may in fact be preferable
for a veto of moderately high efficiency (as it could be operated with
less shielding), for very high-efficiency vetoes there is little, if any,
difference between gadolinium and boron loading in the energy
threshold required for the veto, and hence no difference in the
tolerable rate of external background. We note also that TMB of
acceptable radiopurity for the veto application was identified in
\cite{Borex_proposal}, while some effort would likely be required to
secure Cd or Gd with acceptable levels of contamination.

\section{A Boron-Loaded Scintillator Neutron Veto}
\label{advantages}

The discussion above suggests that a neutron veto efficiency greater
than 99.9\% could be obtained using a 1 meter thick boron-loaded liquid
scintillator and an 11 $\mu$s time window. In actual operation, however,
the veto will surround a detector of finite mass, which leads to the
possibility that neutrons will be captured by components of
the inner detector and hence will not be vetoed.  
 
This effect has been investigated extensively in the context of the
DarkSide-50 experiment. DarkSide-50 is a proposed direct-detection dark
matter experiment based on a 2-phase argon TPC, which will
make use of argon depleted in $^{39}$Ar from recently identified
underground sources \cite{underground_argon}, and which will use the new ultra-low-background
QUPID photon detectors \cite{qupid}. These features, in addition to the use of
a high-efficiency neutron veto, should not only give the 50 kg
experiment a significant physics reach and the ability to make a
convincing dark matter detection claim based on the observation of a
few events, but will also give
the relatively small experiment the potential to
demonstrate directly that a larger detector of similar design could
be operated in a background-free mode for several ton-years.

The DarkSide-50 geometry will essentially consist of a cylindrical
active volume with equal diameter and height, contained in a fused
silica inner vessel, with arrays of QUPIDs on the flat top and bottom
faces. The vessel and QUPIDs are immersed in a (passive) liquid argon
buffer inside a titanium (or low background stainless steel) cryostat\cite{darkside_proposal}. For
the purposes of this discussion of veto efficiency, the important
features of DarkSide-50 are the masses of the
individual detector components. These are listed in Table
\ref{tab:ds_mass}, as implemented in the simulation used here.

\begin{table}[htbp]
\centering
\begin{tabular}{c c c}
\hline\hline
Component & Material & Mass (kg)\\
\hline
Active Region & Depleted Argon & 52.7 \\
Inner Vessel + Photodetectors & Fused Silica & 25.4 \\
Passive Buffer & Depleted Argon & 74.1 \\
Cryostat + Inner Mechanics & Titanium & 78.6\\
\hline\hline
\end{tabular}
\caption{The masses of the different inner components of the
  DarkSide-50 detector, as simulated. Except for the active gas and
  liquid depleted argon, all components are passive absorbers of
  neutrons.}
\label{tab:ds_mass}
\end{table}

In DarkSide-50, which was explicitly designed for
operation in a neutron veto, the materials used in the inner
detector were chosen to minimize the number of neutrons lost to
captures by inactive components, and hence to increase the veto
efficiency. This is in contrast to dark matter detector designs which
attempt to mitigate internal cosmogenic neutrons using (passive)
internal neutron absorbers. We have studied the inclusion of such
passive absorbers, and have found that the increase in veto efficiency
associated with the removal of the absorbers significantly outweighed
the loss of the relatively modest neutron reduction they afforded. 

\subsection{Internal Radiogenic Neutrons}

For the DarkSide-50 detector described above, with the cryostat
surrounded by a 1 meter thick boron-loaded liquid scintillator, radiogenic
neutrons generated in the inner detector, specifically in the
photosensors, produced veto events (as before, at least
40 keV$_{ee}$ of energy deposition within a 1 $\mu$s window was required
to produce a veto signal) with an
efficiency of 99.78$\pm$0.01\%. The inefficiency due to
neutron capture on the inner detector is small, not because neutrons
are not captured by the inner detector (indeed, in this simulation
about 21\% of primary neutrons are captured by inner detector
components), but because the majority of these inner detector captures
result in the production of secondary particles,
particularly gamma-rays, which are subsequently detected by the veto.

\begin{figure}[htbp]
\centering
\scalebox{0.35}{\includegraphics{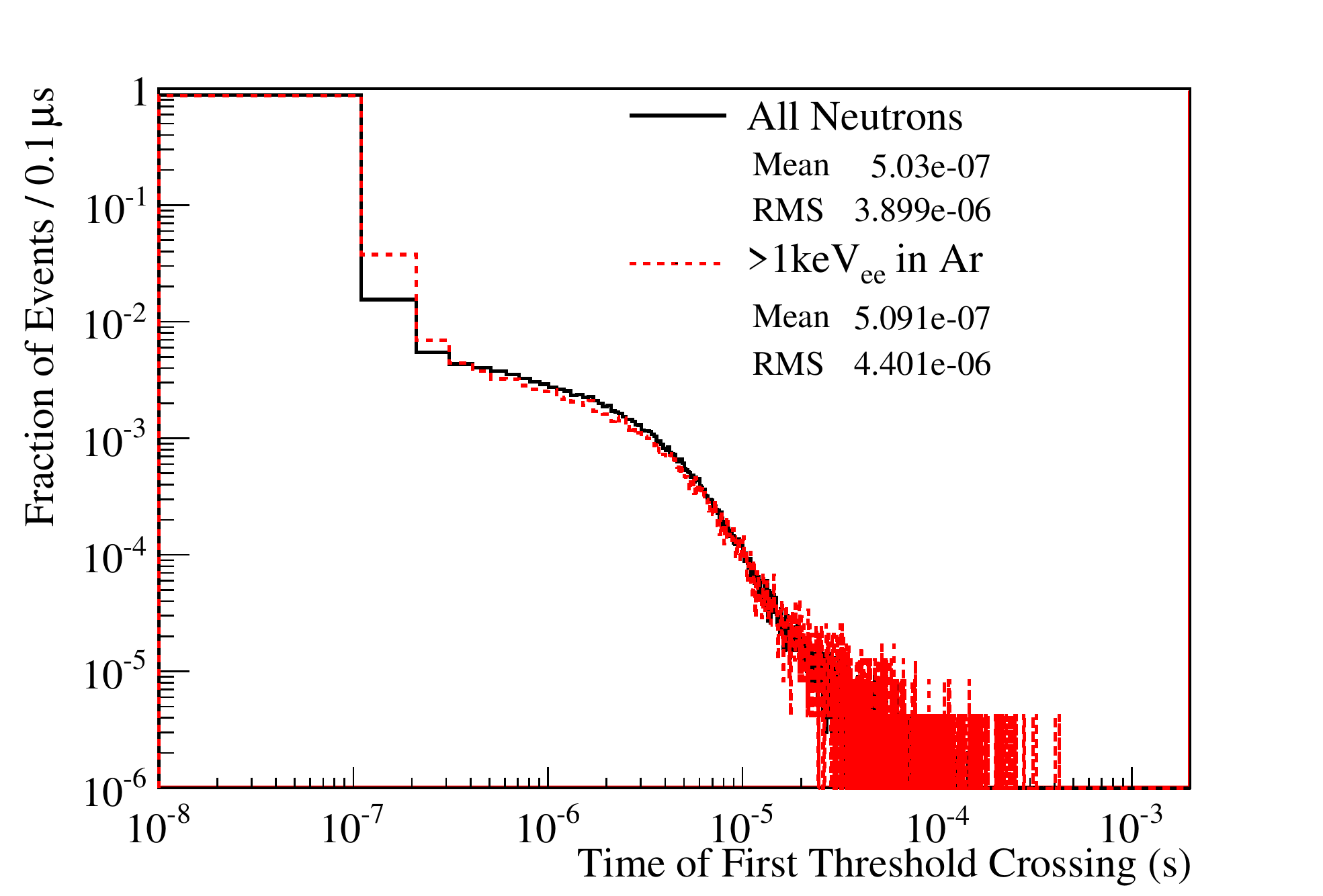}}
\caption{The simulated time distribution of veto events produced by
  radiogenic neutrons from the inner detector. As can be seen, the
  distribution of veto times is quite insensitive to whether or not
  the neutron deposited energy in the argon active volume of the inner detector. The fraction of
  all neutrons vetoed by a prompt recoil was 88\%: for the neutrons
  which deposited some energy in the argon this value is 86\%.}
\label{fig:det_times}
\end{figure}

The potential for neutrons to thermally walk within
the inner detector, the components of which have lower neutron capture
cross sections than the boron-loaded scintillator, for significant
periods of time means that the time distribution of neutron
veto signals is longer with the detector present (see Table
\ref{tab:det_time}) than without (Table \ref{tab:eff_time}), and is no
longer described at longer times by a single exponential (Figure
\ref{fig:det_times}). However, neither the veto times nor the veto efficiency
appear to vary significantly between those neutron events which
deposited energy in the active argon volume and those which did not.

\begin{table}[htpb]
\centering
\begin{tabular}{c c}
\hline\hline
Detection Efficiency & Time Required ($\mu$s)\\
\hline
70\% &  0.08 \\
90\% & 0.37\\
95\% & 2.3\\
98\% & 5.5\\
99\% & 9.3\\
99.5\% & 21.5\\
99.8\% & 57.7\\
\hline\hline
\end{tabular}
\caption{The time interval required after neutron production to
  contain the veto signals with
  different probabilities. The increase in the neutron capture time
  caused by the presence of the inner
  detector can be seen by comparing these
  values with those in in Table \ref{tab:eff_time}.}
\label{tab:det_time}
\end{table}

The efficiency of the neutron veto might also be affected by the
feedthroughs necessary to supply cryogenic cooling to the inner
detector and to operate the
photodetectors. To investigate this potential effect, a 10 cm diameter
air-filled trunk was added to the simulation. The trunk passed through
each of the inner vessel components (other than the active argon), and
then out through the neutron veto. In the center of the veto, a 60 cm long right-angle
``dog-leg'' was added to the feedthrough, to eliminate straight paths of neutron escape. With the feedthrough in place, radiogenic neutrons
generated in the photodetectors produced veto events with
99.75$\pm$0.02\% efficiency, with no apparent effect on the time
distribution of the veto events. Thus, it seems that the necessity of
having a passive feedthrough within the neutron veto will not have a
significant effect on the efficiency with which the neutron veto
rejects internal radiogenic neutrons.

In all, then, our simulations suggest that, with a 60$\mu$s veto window, veto efficiencies higher
than 99.7\% could be obtained with a 1 meter thick neutron veto of
boron-loaded scintillator around the DarkSide-50 detector, even after
including the effects of
neutron capture on the inner detector and a 10 cm diameter
feedthrough in the veto. Such a veto could be expected to provide
similar performance when used with other detectors of similar mass, provided
that some care is taken to minimize neutron loss in the inner
detector. 

\subsection{External Radiogenic Neutrons}

In addition to its high detection efficiency for internally produced
radiogenic neutrons, a boron-loaded scintillator veto is also an
excellent shield
against external radiogenic neutrons. Our simulations suggest that the
fraction of external radiogenic neutrons that will penetrate
the 1 meter neutron veto and produce a recoil in the active volume of
DarkSide-50 is $\lesssim$1x10$^{-7}$. It is expected that the residual external radiogenic events will
be vetoed with an efficiency at least equal to the veto efficiency for
internal radiogenic neutrons, giving an overall reduction in the rate
of external radiogenic backgrounds of more than 10$^9$, although we
have not generated sufficient Monte Carlo
statistics to confirm this.  

\subsection{Cosmogenic Neutrons}

As dark matter experiments continue to increase in sensitivity,
cosmogenic neutron backgrounds become a dominant background, even at
relatively deep sites. These high-energy neutrons are quite
penetrating, and are hence difficult to shield. Nonetheless, with
relatively large shields, which are themselves instrumented as muon
veto systems to reject cosmogenic backgrounds originating from within
the shielding itself, cosmogenic backgrounds can be significantly
reduced. 

A liquid scintillator is ideal for use as a shield against external
cosmogenic neutrons\footnote{Note that the efficiency of the scintillator in vetoing
those muons which pass through the scintillator directly is
sufficiently high that internal cosmogenic backgrounds are strongly
sub-dominant to the external cosmogenics, and as a result internal
cosmogenics are not considered here.}, as the
scintillator has the potential to detect the cosmogenic neutrons,
rather than simply attenuating them. This means that the mean free
path (Table \ref{tab:mfps}), rather than the attenuation length (Table
\ref{tab:last_recoils}), gives the appropriate scale for the reduction
of the flux of external cosmogenic neutrons by a scintillator
veto. In addition, the scintillator veto has the potential to detect
cosmogenic neutrons both before and after they have interacted in
the inner detector, as opposed to attenuation-based shielding which is
effective on the incident particle only. Both the change from
attenuation length to mean free path as the effective metric and the
ability to detect the recoiling as well as the incident neutrons 
increase the effectiveness of the scintillator veto by approximately a
factor of two relative to water shielding, meaning that, in aggregate, 1 meter of scintillator veto is about
as effective as 4 meters of water shielding in reducing backgrounds due to external
cosmogenic neutrons.

\begin{table}[htpb]
\centering
\begin{tabular}{c c c c}
\hline\hline
Neutron Energy & MDR in Water & MDR in Pure & MDR in Loaded\\
(MeV) & (cm) & Scintillator (cm) & Scintillator (cm)\\
\hline
10 & 22.2 & 26.3 & 27.5 \\
20 & 27.4 & 30.8 & 32.5 \\
50 & 54.8 & 60.5 & 62.7 \\
100 & 92.0 & 98.5 & 102.9 \\
200 & 145.7 & 168.2 & 170.4 \\
\hline\hline
\end{tabular}
\caption{The mean position of the most distant nuclear recoil
  (``MDR'') produced
  by neutrons of different energies (or any of their daughter
  particles). This metric provides an ''effective attenuation length''
  appropriate for use in estimating neutron shielding necessary for
  direct-detection dark matter experiments.}
\label{tab:last_recoils}
\end{table}

In order to obtain a quantitative estimate of the ability of the scintillator veto to reduce external
cosmogenic neutrons, these were added to the DarkSide-50
simulation. The energy spectrum used was the parameterized spectrum
given in \cite{Mei_and_Hime} for cosmogenic neutron production at
a depth of 3.1 km.w.e. (the effective depth of LNGS). Only primary
neutrons with energies greater than 12 MeV were simulated, as the lower
energy neutrons are unlikely to penetrate the veto. We note that
the energy spectrum of cosmogenic neutrons emerging from the rock
walls will differ from the production spectrum somewhat due to rock
attenuation - this effect was not considered here. Also, by generating
single cosmogenic neutrons as the primary particles in our
simulations, we neglect the possibility that other particles from the
primary muon shower could enter the veto: this should cause us to
underestimate, perhaps significantly, the efficiency of the veto against external cosmogenic neutrons. 

Even in these very conservative simulations, however, a 1 meter thick scintillator veto reduced the number of recoil events in the active
argon detector produced by cosmogenic neutrons with initial energies
greater than 12 MeV by about a factor of 40 (the number of recoil events were
reduced by $\sim$25\% from attenuation, with $\sim$97\% of
the remainder being vetoed) compared to the rate with no neutron shielding. We note that
from the perspective of external cosmogenic neutrons,
the factor of 40 reduction  is equivalent to increasing
the depth of the experiment by more than 2 km.w.e.

For interest, we have simulated the suppression in external cosmogenic
neutrons possible with thicker scintillator vetoes. As can be seen in
Table \ref{tab:cosmo}, cosmogenic neutron suppression continues to increase
as the veto thickness increases.

\begin{table}[htbp]
\centering
\begin{tabular}{c c}
\hline\hline
Veto Thickness & Relative Recoil Rate in\\
(m) & DarkSide-50\\
\hline
0 & 1.0 \\
1 & (2.7$\pm$0.4)x10$^{-2}$\\
2 & (2.4$\pm$0.6)x10$^{-3}$\\
3 & (5$\pm$2)x10$^{-4}$\\
\hline\hline
\end{tabular}
\caption{The factor by which the rate of external cosmogenic
  backgrounds is reduced (through a combination of attenuation and
  direct vetoing) by neutron vetoes of different thicknesses.}
\label{tab:cosmo}
\end{table}

\section{Veto Trigger Rates and Veto-Induced Dead Time}
\label{backgrounds}

The effective dead time produced by the veto in the inner detector
depends on the event rate in the veto and the veto time window. For a
60 $\mu$s veto window, which from Table \ref{tab:det_time} is necessary
to maintain a very high veto efficiency, a 1\% dead time would be
produced by a veto rate of 168 Hz, while a 10\% dead time would be
produced by a 1756 Hz veto rate.

The background rate in the veto will be due to external backgrounds, intrinsic
backgrounds in the scintillator itself, background decays in the PMTs,
and backgrounds from the inner detector. The latter will presumably be
negligible on the scale of 100 Hz, while the intrinsic rate in
the scintillator will be dominated by $^{14}$C,\footnote{The
  reduction of U and Th backgrounds by distillation was
  demonstrated in \cite{Borex_proposal} to be even more effective in
  TMB than in pseudocumene, so, as in
  \cite{borexino_7be}, the rates of these ``other''
  backgrounds should be small compared to $^{14}$C.} which, at the
10$^{-18}$ $^{14}$C/$^{12}$C ratio found in petrochemically-derived
scintillator \cite{Borex_proposal}, would have a rate in the 11.5T
veto of 2-3 Hz. 

While the background rate from PMT activity and external backgrounds will
depend on the construction and location of the detector, it is
possible to produce reasonable estimates of what these background
rates could be. The total background decay rate in ``low-background'' 8'' PMTs varies
in the range of about 1-6 Hz, depending on the manufacturer. With 80 PMTs (see
Section \ref{optics}), the background rate would likely be 100-500 Hz. With the PMTs
mounted so that their faces are flush with the outer wall of the veto
(as assumed in the optical simulations), perhaps half of this activity
would be detected by the veto.

External backgrounds will be dominated by gamma-rays produced in the
rock surrounding the experiment. A reasonable estimate for the
activity in ``typical'' rock is 15 Bq/kg $^{238}$U + $^{232}$Th
activity, and 300 Bq/kg $^{40}$K activity (there is, of course,
significant variation in these activities between different sites). These
correspond to fluxes of $\sim$2500 $\gamma/m^2/s$ for the 2.2 and 2.6
MeV U and Th chain gamma-rays, and $\sim$42,000 $\gamma/m^2/s$ for the
1.4 MeV gamma from K.

Taking DarkSide-50 as an example, the surface area of the veto will be about 31m$^2$,
so for these levels of external activity we would expect $\sim$80,000
Hz of incident 2.6~MeV photons, and $\sim$1.3x10$^6$~Hz of 1.4~MeV
gammas\footnote{We actually plan to deploy DarkSide-50 within the
  CTF water tank, an 11~m diameter x 10~m high water tank at LNGS,
  which will significantly attenuate external backgrounds. Here we
  discuss a solution which might be more generally useful in other applications of a neutron veto.}. As these rates are too high for effective veto operation, the
gamma-rays must be attenuated. We have explored relatively inexpensive
scrap steel from de-commissioned ships as a passive shield. The
simulated attenuation factors for 2.6 and 1.4 MeV gammas as a function of steel
thickness are given in Table \ref{tab:gamma_att}. For the rate of
external gamma-rays described above, about 25 cm of steel would be
sufficient to reduce the rate of external gamma-ray interactions in
the veto to $\sim$150 Hz. It should be noted that at some underground
sites with lower gamma background rates, as little as half this
passive shielding would be required.

\begin{table}[htbp]
\centering
\begin{tabular}{c c c}
\hline\hline
Steel Thickness & Survival Fraction & Survival Fraction\\
(cm) & for 2.6 MeV gammas & for 1.4 MeV gammas\\
\hline
0 & 1.0 & 1.0 \\
5 & 0.27 & 0.18\\
10 & 0.061 & 0.027\\
15 & 0.013 & 3.9x10$^{-3}$\\
20 & 3.0x10$^{-3}$ & 5.9x10$^{-4}$\\ 
25 & 6.4x10$^{-4}$ & 7.5x10$^{-5}$\\
30 & 1.8x10$^{-4}$ & 1.4x10$^{-5}$\\
35 & 1.7x10$^{-5}$ & 1.8x10$^{-6}$\\  
\hline\hline
\end{tabular}
\caption{The effective attenuation of 2.6 and 1.4 MeV
  gamma-rays in layers of steel of different thicknesses, as simulated using Geant4. The attenuation factor is calculated based on the number
of events produced in the neutron veto by external gamma-rays of the
relevant energies with different thicknesses of steel shielding.}
\label{tab:gamma_att}
\end{table}

An additional  source of radioactive background would then be the steel
itself. Measurements of the ship-breaking steel indicate that the K
activity is less than 13 mBq/kg, while the U and Th activity totals less
than about 2.5 mBq/kg. At these rates (and conservatively taking all U
and Th decays to produce 2.6 MeV gamma-rays), the simulation predicts
a rate of about 65 Hz in the veto from activity in the steel.

A final source of veto-induced dead time in the inner detector will be
random triggers produced by dark rate pile-up. The 80 8'' PMTs that
we assume will be used to instrument the veto can be reasonably assumed to have a 1 kHz dark
rate. Requiring a coincidence between at least three PMTs within a
1 $\mu$s veto trigger window, and assuming that the dark hits are
randomly distributed in time, the rate of random veto triggers can be
calculated to be 80 Hz. 

Although the trigger rates induced in the veto by the different
sources discussed above will vary depending on the
location of the detector and the construction of the veto, we have
nonetheless produced reasonable estimates for the rates in a ``typical''
installation. As shown in Table \ref{tab:total_vet}, this ``typical''
rate would be quite acceptable, with the inner detector incurring a
dead time of about 3\% from the veto. 

\begin{table}[htbp]
\centering
\begin{tabular}{l c}
\hline\hline
Background Source & Veto Rate (Hz)\\
\hline\hline
Inner Detector & $<$1 \\
Scintillator Background &  3 \\
PMTs & 200 \\
External Backgrounds with 25 cm steel & 150 \\
Steel Backgrounds & 65 \\
Random Veto Triggers (1 kHz dark rate) & 80\\
\hline
{\bf Total Veto Rate} & 498\\
\hline\hline
\end{tabular}
\caption{The estimated rate of background events in the veto for a
  ``typical'' underground installation, with 25 cm of passive steel
  shielding. This total veto rate corresponds to a dead time of about
  3\% in the argon detector.}
\label{tab:total_vet}
\end{table}

\section{Veto Optical Efficiency}
\label{optics}

\begin{figure}[htbp]
\centering
\scalebox{0.35}{\includegraphics{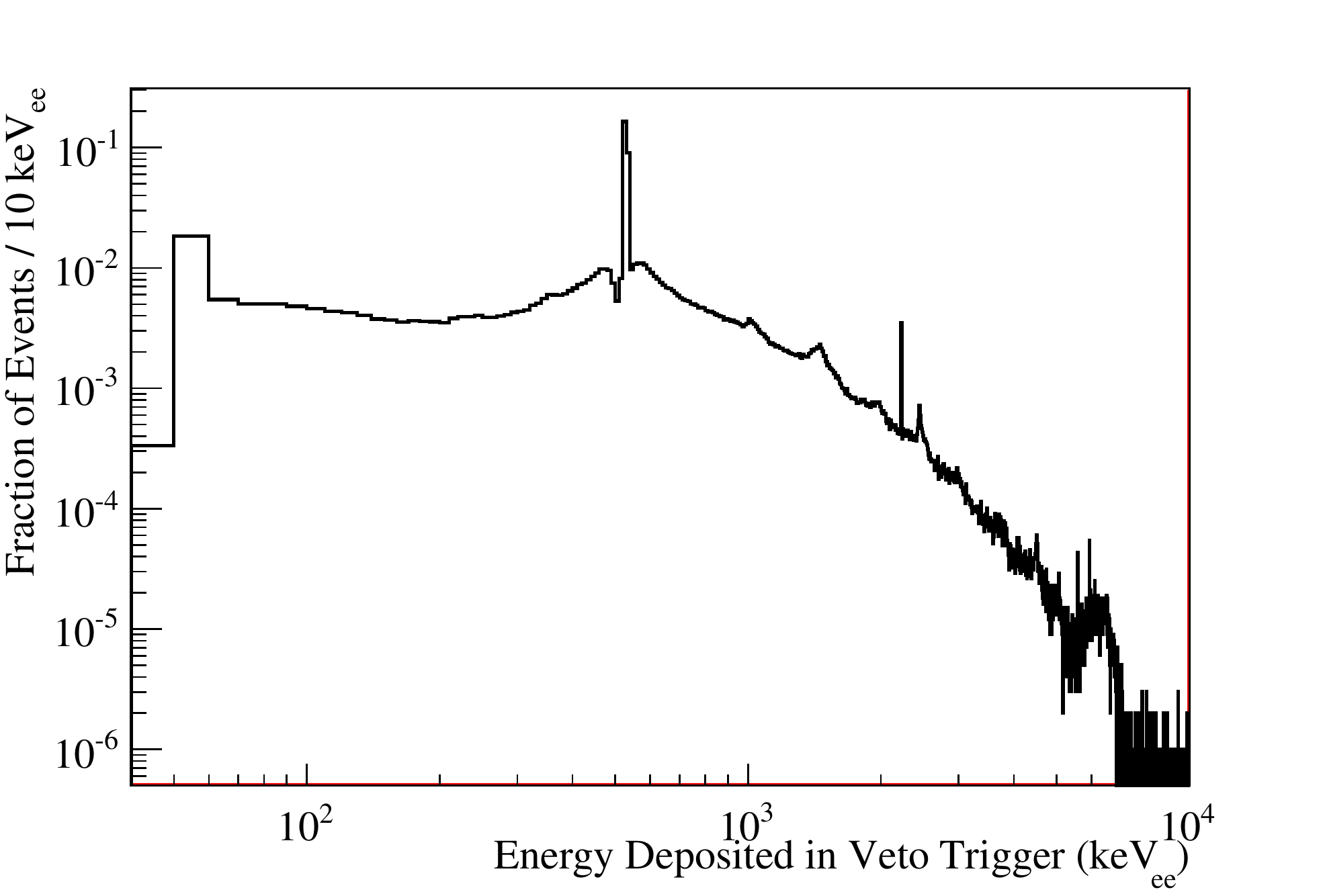}}
\caption{The energy spectrum of veto triggers generated by radiogenic
  neutrons. In each case the energy is deposited within a 1 $\mu$s
  window; if more than one veto trigger was generated by a neutron
  event, only the earliest is plotted. The DarkSide-50 inner detector
  was present in this simulation - gamma lines from neutron capture on
  the inner detector components, as well as energy depositions from
  neutron scattering and
  capture within the veto itself can be seen in the spectrum.}
\label{fig:espec}
\end{figure}

The 40 keV$_{ee}$ detection threshold in the veto assumed in the
earlier analysis was chosen so as to include the
($\approx$50 keV$_{ee}$) energy deposition by
the recoil daughters after neutron capture on $^{10}$B. As can be seen
in Figure \ref{fig:espec}, fewer than 5\% of veto triggers deposit
less than 100 keV$_{ee}$. Nevertheless, it is important that the veto
be able to reliably detect events at the threshold level in order to
maintain the very high veto efficiency.

The chief concern in detecting such low energy events, of course, is
the collection of a sufficient number of photons from each event to
identify the neutron capture. We have performed a baseline study of
the optical efficiency of the scintillator veto if it were instrumented
with 80 standard 8'' PMTs evenly distributed over the outer (curved)
side of the veto cylinder, with no PMTs on the top and bottom. A
rather crude optical simulation, independent of the Geant4 based
physics simulation, was written to study the light collection
efficiency of the veto in such a configuration. The model assumed:
\begin{enumerate} 
\item A 20\% quantum efficiency for the 8'' PMTs. Photons that hit a
  PMT and were
  not detected were assumed to have a 20\% probability of being
  reflected back into the veto.
\item A 95\% reflection probability on all surfaces other than the
  PMTs (i.e. the inside walls of the veto tank and the outside of the
  DarkSide-50 cryostat).
\item An average scintillator light output of 6,000
  photons/MeV$_{ee}$. The statistical fluctuation in the light output
  was assumed to be Gaussian, with a width of $\sqrt{N}$ .
\item A 5 m optical absorption length in the scintillator, with no probability
  of re-emission.
\item A 2 m optical scattering length. 
\end{enumerate} 
Each of these assumptions is extremely conservative:
\begin{enumerate}
\item 8'' PMTs with quantum efficiencies $>$30\% are currently
  commercially available.
\item Both Spectralon PTFE and 3M Vikuiti foils, which are commercially
  available, have reflectances in excess of 95\%
  for the relevant wavelengths.
\item Carefully prepared pseudocumene scintillator has typical light
  output of $\sim$12,000 photons/MeV$_{ee}$ \cite{borex_light_yield}, and we have assumed that
  light output scales with TMB dilution. It is known, however, that the light output of
  scintillator with an optically inert dilutant, like TMB, typically
  decreases more slowly than the concentration\footnote{For example,
    the KamLAND scintillator, which consists of pseudocumene diluted
    to 20\% in dodecane, has a light output at least $\sim$80\% that of
    pure PC \cite{kamland_light_yield}.}. 80\% (rather than 50\%) TMB-loaded scintillator
  with a light
  output of 6000 photons/MeV was reported in \cite{Borex_proposal}.
\item Scattering and absorption lengths in TMB loaded scintillator
  greater than 10 m are reported in \cite{Borex_proposal}. 
\end{enumerate}

\begin{figure}[htbp]
\centering
\subfigure[]{\scalebox{0.35}{\includegraphics{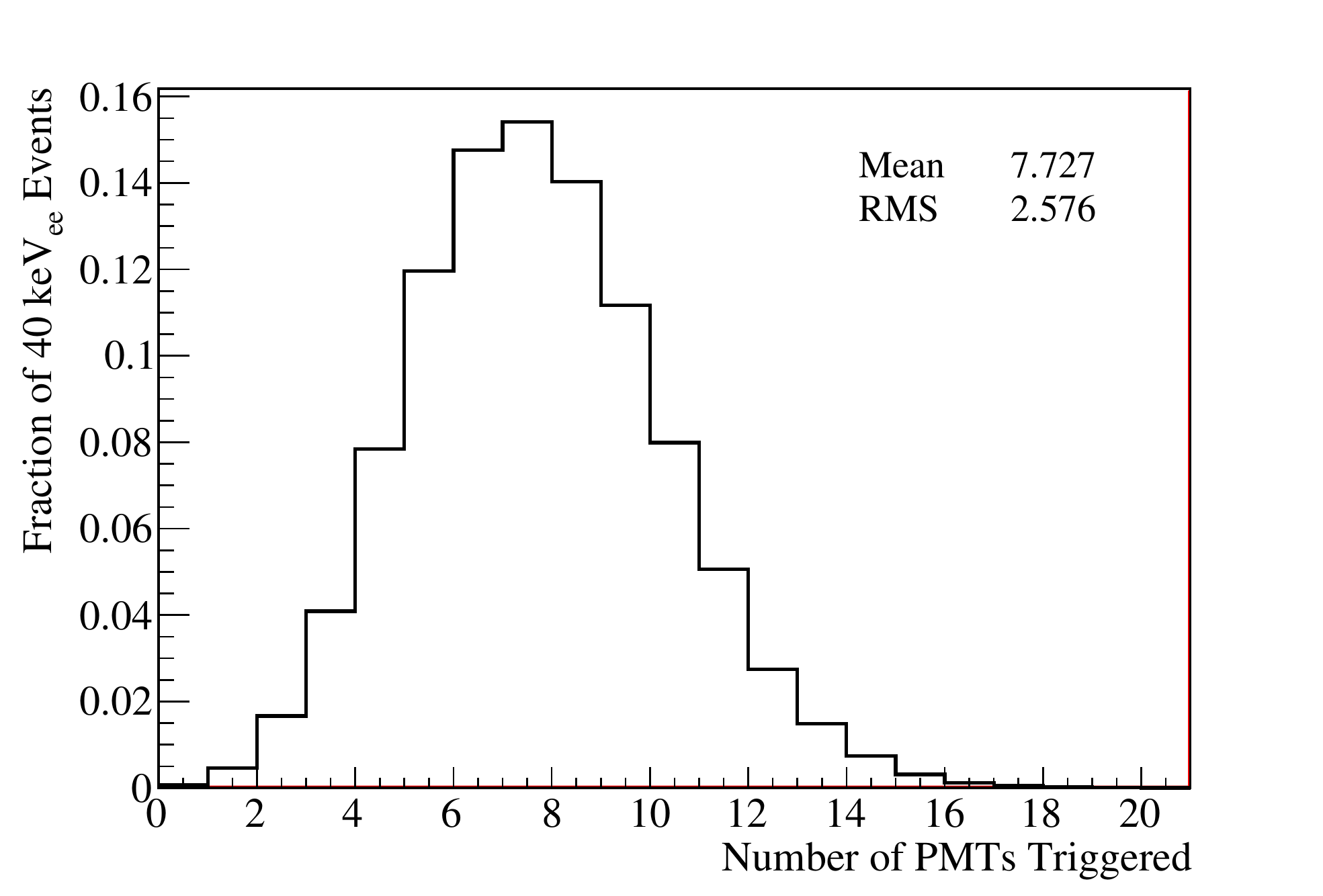}}}
\subfigure[]{\scalebox{0.35}{\includegraphics{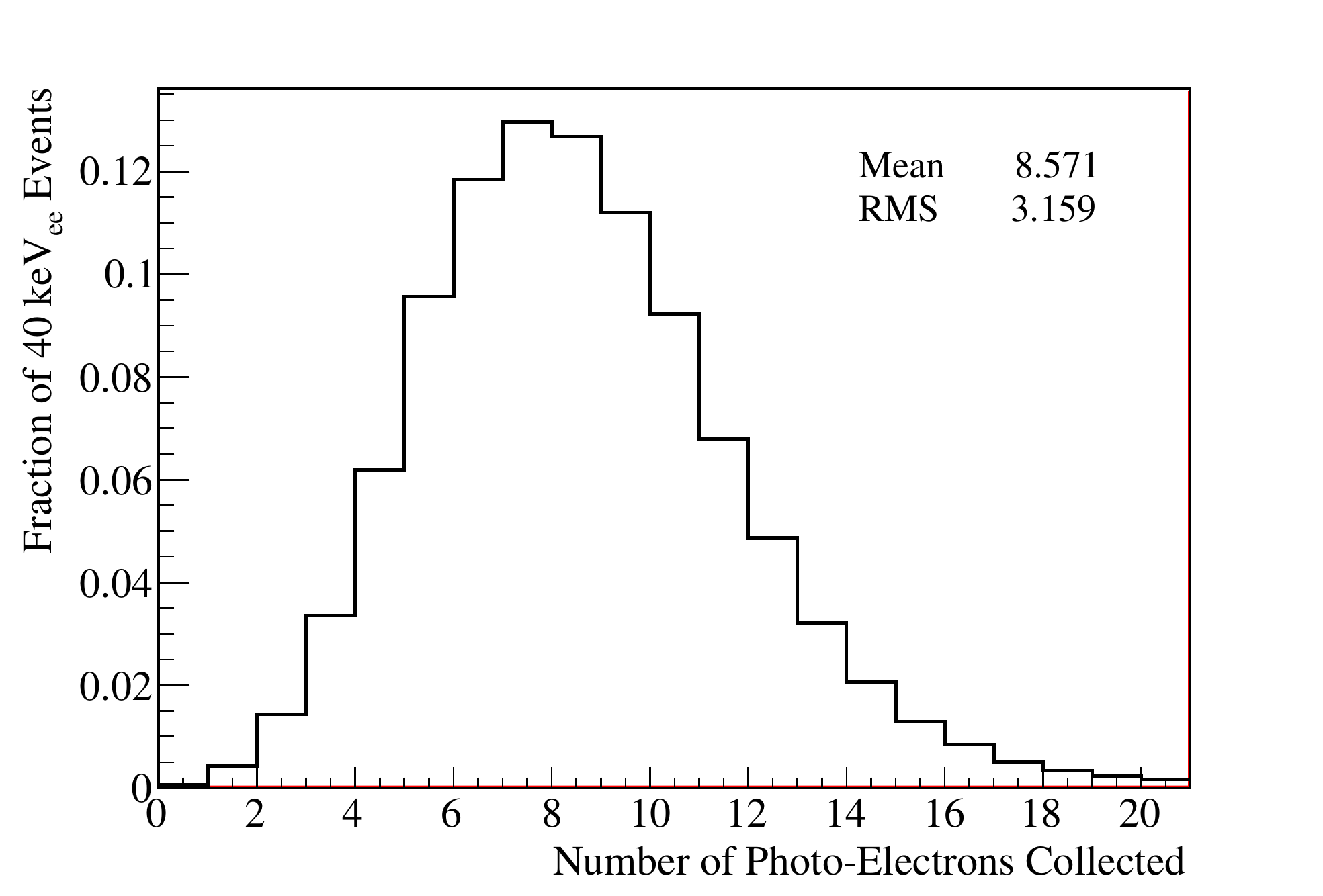}}}
\caption{The distributions of (a) the number of PMTs which
  detected at least one photoelectron and (b) the total number of collected
  photoelectrons in simulations of 40 keV$_{ee}$ events uniformly distributed
  through the neutron veto.}
\label{fig:light_collection}
\end{figure}

Figure \ref{fig:light_collection} shows the simulated distributions of
the total number of collected
photoelectrons and the number of PMTs which detected at least one
photoelectron when events of exactly 40 keV$_{ee}$ were simulated uniformly distributed
throughout the veto volume. As
discussed in Section \ref{backgrounds}, in order to maintain a reasonable rate of random triggers, a
coincidence of 3 PMT hits would be required to trigger a veto
event. As can be seen, even with the very conservative assumptions
about the optical performance of the veto discussed above, in this
simulation about 98\% of 40 keV$_{ee}$
events produce a veto trigger. Given that events below 100 keV$_{ee}$
constitute less than 5\% of the total, the overall fraction of
neutron-induced backgrounds missed due
to veto trigger inefficiency can therefore be expected to be
significantly less than 0.1\%. 

\section{Conclusions}
\label{conclusions}

Neutron vetoes with very high efficiencies can be produced for direct
detection dark matter experiments by surrounding the WIMP detector
with a layer of liquid scintillator. Even after considering neutron
loss in the inner detector and veto inefficiency due to feedthrough
connections, our simulations suggest that a 1 meter thick veto can provide
greater than a 99.5\% efficiency for rejecting background events due
to internal radiogenic
neutrons, while reducing the background from external cosmogenic
neutrons by more than 95\%. Loading $^{10}$B, or
another isotope with a high neutron capture cross section, into the
scintillator makes the veto practical by greatly reducing the neutron
capture time. This reduces the veto window necessary for high
veto efficiency sufficiently so that conventional PMTs can be used in the
veto, and typical underground background rates can be
tolerated with relatively little external shielding.

\section{Acknowledgements}

The authors would like to thank the other members of the DarkSide
collaboration and the members of the CDMS group at FNAL for
interesting and productive discussions about the practical aspects of
the deployment of liquid scintillator neutron vetoes in dark matter
experiments.





\bibliographystyle{model1a-num-names}
\bibliography{bibliography}







\end{document}